\documentclass{emulateapj}

\usepackage{color}
\usepackage{amssymb,amsmath}
\usepackage{graphicx}
\usepackage{epstopdf}

\begin{document}

\title{Interpretation of  the 115 day periodic modulation in the X-ray flux of NGC 5408 X-1}

\author{D. L. Foster$^{1,2,\star}$, P. A. Charles$^{1,3}$, and K. Holley-Bockelmann$^{2,4}$}

\affil{
$^{1}$South African Astronomical Observatory, PO Box 9, Observatory, 7935, South Africa\\
$^{2}$Vanderbilt University, Department of Physics \& Astronomy, Nashville, TN\\
$^{3}$University of Southampton, School of Physics \& Astronomy, Southampton, UK\\
$^{4}$Fisk University, Department of Physics, Nashville, TN}

\altaffiltext{$\star$}{E-mail: deatrick@saao.ac.za (DLF)}

%\author{D. L. Foster\altaffilmark{1}, P. A. Charles\altaffilmark{2}}
%\affil{South African Astronomical Observatory, PO Box 9, Observatory, 7935, South Africa}
%\email{deatrick@saao.ac.za, pac@saao.ac.za}
%
%\and
%
%\author{K. Holley-Bockelmann\altaffilmark{3}}
%\affil{Vanderbilt University, Department of Physics \& Astronomy, 1807 Station B, Nashville, TN 37235}
%\email{k.holley@vanderbilt.edu}
%
%\altaffiltext{1}{Vanderbilt University, Department of Physics \& Astronomy, 1807 Station B, Nashville, TN 37235}
%\altaffiltext{2}{University of Southampton, School of Physics \& Astronomy, Southampton, Hampshire, SO17 1BJ, UK}
%\altaffiltext{3}{Fisk University, Department of Physics, 1000 17$^{th}$ Avenue North, Nashville, TN 37208}

\begin{abstract}
We comment on the recent observation of a 115-day modulation in the X-ray flux of the ultraluminous X-ray source (ULX) NGC 5408 X-1, and in particular, the interpretation of this modulation as the orbital period.  We suggest that this modulation may instead be due to a precessing jet, and is thus {\it superorbital} in nature.  Comparing the properties of this ULX with those of the prototypical micro-quasar SS~433, we argue that NGC 5408 X-1 is very similar to SS~433:  a hyper-accreting stellar-mass black hole in a shorter-period binary. If the analogy holds,  the 115-day modulation is best explained by the still poorly-understood physics of inner-disc/jet precession and a longer observing baseline would be able to reveal an intrinsic phase jitter that is associated with such a precession.
\end{abstract}


\section{Introduction}
The nature of ultra-luminous X-ray sources (ULXs) discovered by {\it Chandra} and {\it XMM-Newton} surveys of nearby galaxies remains highly controversial \citep[e.g.][]{Fabbiano:2006, Zampieri:2009}.  ULXs appear as point-like, non-nuclear X-ray sources characterised by extremely high X-ray luminosities ($L_X \ga 10^{39}$ erg s$^{-1}$) compared to {\it Galactic} X-ray binaries (XRBs) \citep*{Long:1981, Fabbiano:1989, Makishima:2000, Fabbiano:2001, Colbert:2004}.  

Making seemingly reasonable assumptions that ULXs emit radiation isotropically and that they are Eddington limited, one might naturally conclude that ULXs harbour the long-sought intermediate-mass black holes (IMBHs) whose masses are $\sim 10^{2-4}$~M$_{\odot}$ \citep{MandC:2004, Miller:2005, Fabbiano:2006}.  If IMBHs exist, they would be too light to plunge to the centres of their host galaxies via dynamical friction, but would be massive enough to power ULX luminosities.

NGC 5408 X-1 is one of the brightest ULXs, with a well-determined X-ray luminosity of $L_X \approx 2\times10^{40}$erg~s$^{-1}$ in the 0.3--10 keV band \citep{Strohmayer:2009}.  Presently, this source is one of the best IMBH candidates, with mass estimates well in excess of 100~M$_\odot$, but possibly up to 9000~M$_\odot$ \citep*{Kaaret:2003, Strohmayer:2007, Strohmayer:2009, Strohmayer:2009b}.  Such large range of uncertainty in the mass must surely be a cause for concern.  This is particularly true given that a number of studies have suggested that ULXs may merely be stellar-mass black hole binaries (BHBs) that exhibit super-Eddington accretion which can be achieved via beaming of the emission, super-critical mass accretion, or a combination of both \citep*{Begelman:2006, King:2008a, King:2008b, King:2009}.  

As a prime example of the difficulty in using the X-ray luminosity to pin down the black hole mass, consider the case of SS~433, a remarkable Galactic X-ray source and the prototypical micro-quasar~\citep[see][]{Margon:1984}.  SS~433 very likely contains a total mass $\sim$ 40 M$_\odot$, with a $\sim$ 4 M$_\odot$ black hole as the compact accretor \citep*{Blundell:2008, Kubota:2010}, and with the bulk of its energy output mechanically beamed \citep[e.g.][]{Begelman:2006}.  Indeed, supporting evidence for the stellar-mass black hole scenario lies in the relativistic jets which precess with a 162-day period about the binary orbital axis.  In addition, there are well-studied massive outflows due to hyper-Eddington mass transfer \citep{Fabrika:2004, Begelman:2006, Blundell:2008}.  The jet itself has a kinetic luminosity $\ga$ 10$^{39}$~erg~s$^{-1}$, which exceeds the observed $L_{\rm X}$ by a factor of $\sim 10^3$. However, its high orbital inclination obscures the intrinsic X-ray brightness from direct view.  All of this together suggests that if viewed externally and at lower inclination, SS~433 would be bright enough to appear as a ULX \citep{Fabrika:2001}. Under these circumstances it would be miscategorised as an IMBH~\citep{Begelman:2006}.  

Model-independent system mass estimates are possible in Galactic X-ray binaries based on their kinematics \citep[][]{Charles:2006}. In the case of SS~433, the mass of the X-ray emitting accretor plus the accretion disc is $\approx 16 M_\odot$ \citep{Blundell:2008}.  However, such system mass estimates have not yet been possible for any extragalactic ULX.

Recently, a case for an IMBH in NGC 5408 X-1 has been made, citing its principal observed property of an extremely high $L_X$ and the presence of a low-frequency quasi-periodic oscillation (QPO).  The X-ray luminosity alone implies a black hole mass of $\sim$ 100~M$_\odot$ if we assume that the ULX is radiating at the Eddington limit.  Adding weight to the interpretation that this source is an extreme form of high mass X-ray binary, there is a recent discovery of a 115-day modulation in the X-ray flux.  \cite{Strohmayer:2009} interpreted this modulation as the binary orbital period for a system with a $\sim$ 1000 M$_\odot$ black hole for the primary component and a 3--5~M$_\odot$ giant or supergiant star for the secondary component.  

In examining the properties of NGC 5408 X-1, we suggest here that it may be instructive to make further comparisons with SS~433.  Both sources are known to be engulfed in photo-ionised, steep-spectrum radio nebulae of similar size \citep{Margon:1984, Soria:2006, Lang:2007, Poutanen:2007,Kaaret:2009}, strongly favouring the presence of optically-thin synchrotron emission powered by an accreting black hole over X-ray emission from a supernova remnant (SNR).  And both may have formed from HMXBs undergoing mass transfer on a thermal time-scale \citep{King:2000}.  This raises the question of whether NGC 5408 X-1 may be more similar to SS~433 than previously realised.  There now exists an extensive observational database for SS~433, particularly for its precessing jets \citep{Eikenberry:2001}, and we investigate here whether its properties are sufficiently similar to those of NGC 5408 X-1 to consider them both part of the same population.

\section[]{Comparing NGC 5408 X-1 and SS~433}

\subsection{Properties of NGC 5408 X-1}
NGC 5408 is a dwarf irregular galaxy at a distance of 4.8~Mpc \citep{K:2002}.  \citet{Kaaret:2003} discovered radio emission at the position of the ULX NGC 5408 X-1, and they found that the X-ray, optical, and radio fluxes were consistent with beamed emission from a relativistic jet of an accreting stellar-mass black hole (although they could not rule out jet emission from an IMBH.)  However, subsequent radio observations \citep{Lang:2007} showed that, in fact, the radio emission was too extended (1\farcs5 $-$ 2\farcs0, or $R \approx 35-46$~pc) to be associated with relativistic jets and was more likely optically thin synchrotron emission from a surrounding nebula.

Because the ULX is located in a relatively unobscured and uncrowded region, its optical counterpart has been identified \citep[][]{Kaaret:2009}.  It has been shown that NGC 5408 X-1~is confined within a photo-ionised nebula (of size $R \sim$ 30 pc) displaying strong high-excitation emission lines of He~II $\lambda$4686 and [Ne~V] $\lambda$3426.  The optical continuum emission of the counterpart was weak and there were no absorption features present that might be associated with the donor's photosphere, thereby precluding a kinematic study \citep[cf. Prestwich et al. 2007 in the case of IC~10~X-1;][]{Kaaret:2009}.  

A potentially crucial feature of this ULX is the presence of a low frequency ($\simeq$ 10 mHz) quasi-periodic oscillation (QPO) in the soft X-ray band.  \citep{Strohmayer:2007, Strohmayer:2009}.  The physical origin of such QPOs remains controversial \citep[][]{Poutanen:2007}.  However, QPO phenomenology has been studied over the last decade, and it is commonly assumed that the black hole mass is inversely proportional to the QPO frequency at a given value of the power-law spectral index \citep*{TLM:1998, TF:2004, SandT:2009}.  Given this assumption, the observed QPO frequency suggests a black hole mass well in excess of 1000~M$_\odot$, possibly as high as 9000 M$_\odot$~\citep{Strohmayer:2009b}.  If these properties were confirmed, then NGC 5408 X-1 would indeed be a remarkable and extremely important object for black hole population studies.

\begin{table}
 \centering
 \begin{minipage}{90mm}
  \caption{Comparing QPO frequency, P$_{sup}$, and M$_X$ in BHBs.}
  \begin{tabular}{@{}lccc@{}}
  \hline
Source & P$_{sup}$ [d] & $f_{QPO}$ [Hz] & M$_X$ [M$_\odot$]\\
\hline
NGC 5408 X-1 & 115.5 $\pm$ 4$^a$ & 0.010$^a$ & ---\\
SS~433 & 162.375 $\pm$ 0.011$^b$ & 0.100$^c$ & 4.3 $\pm$ 0.6$^d$\\
GRS 1915+105 & 590 $\pm$ 40$^e$ & 0.001--67$^f$ & 14 $\pm$ 4$^g$\\
GRO J1655-40 & $\simeq$ 3$^i$ & 0.1--450$^{j,k,m}$ & 6.3 $\pm$ 0.5$^n$\\
Cygnus X-1 & $\simeq$ 300$^p$ & 0.040--0.070$^q$ & 21 $\pm$ 8$^p$\\
\hline
\hline

\end{tabular}

\parbox[s]{8.5cm}{
\footnotesize{~$^a$\cite{Strohmayer:2009}.  $^b$\cite{Eikenberry:2001}.  $^c$\cite{Kotani:2006}.  $^d$\cite{Kubota:2010}.  $^e$\cite*{Rau:2003}.  $^f$\cite*{Morgan:1997}.  $^g$\cite{Greiner:2001}.  $^i$\cite{Hjellming:1995}.  $^j$\cite{Remillard:1999}.  $^k$\cite{Strohmayer:2001}.  $^m$\cite{Remillard:2002}.  $^n$\cite{Greene:2001}.  $^p$\cite*{Rico:2008}.  $^q$\cite{Vikhlinin:1994}.  }
}

\end{minipage}
\end{table}

\subsection{Properties of SS~433}

A twenty-year baseline of many optical photometric and spectroscopic campaigns has revealed multiple periodicities in SS~433:  a well-
established orbital period of $P_{\rm orb}$ = 13.07~d \citep*{Crampton:1980}, a precessional period $P_{\rm prec}$ = 162.375~d \citep{Eikenberry:2001}, and a nutation period $P_{\rm nu}$ = 6.3~d \citep{Mazeh:1980, Katz:1982} due to a periodic torque produced by the secondary component 
which does not affect the mean precession rate but does produce an instantaneous oscillation in the precession with a period of about one-half the 
orbital period  \citep{Katz:1982, Bate:2000}.

SS~433 has an observed X-ray luminosity of $L \sim 10^{36}$ erg s$^{-1}$, but as an eclipsing binary, it has a high inclination and fits the 
description of a classic {\it accretion disc corona} (ADC) source \citep*{Watson:1986, Frank:1987}; the obscuration by the disc implies that the 
intrinsic $L_{\rm X}$ is a factor 10$^{2-3}$ higher.  The observed kinetic luminosity in the jets is in the range $L_k \sim 10^{39-41}$ erg s$^{-1}$ 
\citep[][]{Margon:1984, Fabrika:2004}.    As pointed out in \cite{King:2008b}, this kinetic luminosity is already larger (possibly very much so) than the 
Eddington luminosity for a $\sim$ 10~M$_\odot$ black hole.  The mass ejection rate from the jets is $\dot{M}_{\rm jets} \ga 5\times10^{-7} M_\odot$ 
yr$^{-1}$ \citep{Begelman:2006}.

SS~433 is located at the centre of the supernova remnant W50 whose size is $ \approx$ 1.5$\degr$~across its largest dimension, corresponding to 
$\approx$ 64 pc at its distance of 5.5 kpc \citep[][]{Margon:1984}.  It is worth noting that this is comparable in extent to that of the NGC 5408 X-1 
optical nebula.  W50 has long been known to exhibit ``ears'' that align with the SS~433 jets, perhaps caused by the strong outflowing wind velocity 
of $\approx$ 1500~km~s$^{-1}$ \citep{Fabrika:2004}.  These winds may be driven by hyper-Eddington accretion onto the black hole at $\approx 5000 \dot{M}_{\rm Edd}$ (Begelman et al. 2006;  see also Fig. 9 of Shakura \& Sunyaev 1973).  Such a hyper-accreting object will generate a 
mechanically beamed accretion luminosity \citep[][]{Begelman:2006, King:2008a}.  Consequently, most of the light is geometrically collimated into a cone along the outflow axis with a solid angle $\Omega$ that cannot be estimated {\it a priori}, since its inverse determines the apparent luminosity.  For consistency with the number of progenitor high mass X-ray binaries in the Milky Way, it is assumed that $\Omega/4\pi \ga 0.1$ for SS~433~\citep{Begelman:2006, Fabrika:2007, Poutanen:2007, King:2008a, King:2008b, King:2009}.

\section[]{The nature of the 115-day periodic modulation in NGC 5408 X-1}
A 115\fd5 modulation was detected in the X-ray flux of NGC 5408 X-1 with the {\it Swift}/X-ray Telescope and was interpreted as being orbital in 
origin \citep[][]{Strohmayer:2009}.  Since the stability of this modulation has not yet been established, this conclusion was largely based on the 
assertion that 115 days is shorter than all other known superorbital periods for stellar mass BHBs in the Galaxy.  However, it is not substantially less 
than the well-established superorbital precession period for SS~433.

 In fact, the question of whether there is significant overlap in the orbital timescales of black holes and neutron stars remains unanswered to date.  Superorbital periods in neutron stars are known to range from tens to hundreds of days \citep{Wen:2006, Charles:2008}---in some cases even longer than the 115 days in NGC 5408 X-1 \citep{Charles:2008, Kotze:2010}.  Taking into account that there are considerably more known neutron star binaries than BHBs, that there are a comparatively small number of known, persistently bright black hole candidates, and that there are {\it several} mechanisms to explain these periodicities \citep*[i.e. the Kozai mechanism, radiation-driven disc warping, etc.;][]{Kozai:1962, Maloney:1996, Maloney:1998}, the need for a comprehensive statistical analysis of the orbital timescales in neutron stars and black holes is apparent.

So far, the modulation period in NGC 5408 X-1 has only been determined to $\pm$ 4~days as a result of the small number of cycles sampled.  We 
note that if the 115 d modulation is orbital, then longer observing baselines will define this more and more precisely, whereas a superorbital 
modulation (such as for SS~433) will suffer an intrinsic phase jitter.  At this time, the observations cannot distinguish between the two interpretations 
and we propose here that the modulation is instead superorbital in origin and hence similar to SS~433.

\section{Discussion}

\subsection{Mass estimates from QPOs}
For Galactic BHBs with high Eddington ratios ($L/L_{\rm Edd}$), there exist kinematic data that constrain their masses to be $\sim$ 10--15 M$_\odot$ \citep[e.g. GRS 1915+105,][]{Greiner:2001}.  However, for ULXs there is no such dynamical information due to their extreme optical faintness and lack of direct detection of the donor.  Therefore, we must use indirect methods.  

Under the assumption that they are signatures of Keplerian orbits of the material in the inner accretion flow, low-frequency QPOs have been used to infer the masses of the primary components of BHBs.  This line of enquiry seems promising in light of more recent work \citep{SandT:2009}, although the techniques involved have thus far only been applied to a small number of stellar-mass black holes (M$_{BH}$ $\sim$ 5$-$15 M$_\odot$), so the range of masses sampled is low.

NGC 5408 X-1 is a ULX whose observed properties suggest two different masses for its compact accretor.  The $L_{\rm X}$ alone implies a $\sim$ 100~M$_\odot$ black hole if we assume a high Eddington ratio.  Yet, this ULX exhibits a low-frequency QPO in its soft flux, suggesting the presence of an IMBH of mass 2000$-$5000 M$_\odot$ \citep{Strohmayer:2009b}.  However, similar low-frequency oscillations have been observed in Galactic binaries that are well-established {\it stellar-mass} binary systems (see Table 1).  For example, GRS 1915+105 exhibits low-frequency QPOs $\sim$ 0.001$-$10~Hz \citep*{Morgan:1997} and Cygnus X-1 has low-frequency QPOs $\sim$ 0.04$-$0.07~Hz \citep{Vikhlinin:1994}.  Furthermore, the sample of stellar mass BHBs used to determine the scaling relationship (which is based on the range of the best-fitting power-law slopes in their reference sample of sources) for NGC 5408 X-1 in \citet{Strohmayer:2009b} uses systems with known stellar-black-hole masses whose observed Eddington fraction is much higher than would be the case for NGC 5408 X-1 if its black hole has a mass of 5000~M$_\odot$ (as predicted in Strohmayer \& Mushotzky 2009) and therefore $L_{\rm Edd} \ga 10^{42}$ erg s$^{-1}$.  The observed $L{\rm_X} \approx 2 \times 10^{40}$ erg s$^{-1}$ is only $\sim$ 2\% $L_{\rm Edd}$ and this may not be reasonable given the environment of this ULX and the apparently steady flux level.  Therefore, one should exercise caution when using the presence of low-frequency QPOs to infer black-hole masses.

\subsection{Mass Transfer from the Donor}
Two distinct effects of super-Eddington mass transfer can cause the X-ray luminosity to exceed the Eddington limit.  First, the disc accretion luminosity has an additional, logarithmic component due to the inner transition region:
% Accretion Luminosity
\begin{align}
L_{\rm{acc}} \simeq L_{\rm{Edd}} \left[1+\rm{ln}\left(\frac{\dot{M}}{\dot{M}_{Edd}}\right)\right]
\end{align}
which could enhance the luminosity by as much as 5$-$10 times $L_{\rm Edd}$ \citep[][]{Poutanen:2007}.  Secondly, the luminosity is modified by mechanical (i.e. geometrical, non-relativistic) beaming, resulting in collimation by a factor $b$ (typically~$\ga$~0.1):
% Apparent bolometric luminosity
\begin{align}
L \simeq \frac{L_{\rm{Edd}}}{b} \left[1+\rm{ln}\left(\frac{\dot{M}}{\dot{M}_{Edd}}\right)\right]
\end{align}
\citep*{King:2002, Rappaport:2005, Poutanen:2007, King:2008a, King:2008b}. Hence it should not be considered surprising that X-ray emitting binaries can exceed $L_{\rm{Edd}}$ by as much as an order of magnitude in certain circumstances.

A key point to consider is that for SS~433, a system with mass ratio {\it M$_2 /$M$_X$} $>$ 1, the inferred mass transfer rate ($\sim$ 10$^{-4}$~M$_\odot$~yr$^{-1}$) requires that it is in a very short-lived phase \citep{Begelman:2006}.  The clear implications are that either such systems are very rare, or their mass transfer rates are highly variable and/or the systems are very young, being located in star-forming regions.

\subsection{Is the 115-day modulation due to disk/jet precession?}

The observed modulation in the X-ray flux of NGC 5408 X-1 is $\approx 13-24$\%, depending on energy.  Roughly the same levels of modulation were observed in SS~433 \citep[][]{Gies:2002}.  These energy-dependent modulations in both SS~433 and NGC 5408 X-1 could be a common feature of such systems, and the spectral effects are perhaps simply related to the orientation of the precessing discs relative to our line of sight.  

\begin{figure}
\includegraphics[width=8cm]{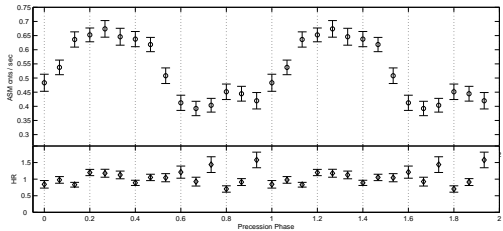}
\caption{{\it Top panel}:  RXTE/ASM count rate (vertical axis, in cts s$^{-1}$) for SS~433, plotted against precession phase (horizontal axis).  {\it Bottom panel}:  ASM hardness ratio, defined as C-band~/~(A-band + B-band), versus precession phase. The precession period is 162 days \citep{Eikenberry:2001}.}
\label{lightcurvehard}
\end{figure}

Figure \ref{lightcurvehard} illustrates the relationship between the modulation and hardness ratio for SS~433.  The relative flatness of the hardness ratio for SS~433 compared with that of NGC 5408 X-1 \citep[cf. Figs. 3 and 4 of][]{Strohmayer:2009} may be accounted for by the difference in viewing angle; that is, SS~433 is an ADC source whose spectrum is modified by its obscuring corona but the same effect is less pronounced in NGC 5408 X-1 because of its lower inclination.  Notably, the effect of absorption and emission of X-rays by a partially ionised wind from a companion star was suggested as a link between the energy-dependent amplitudes of the 115-day modulation and the orbital period of the system \citep{Strohmayer:2009}; however, the same effect should be present in the case of SS~433 but is not.  Also, the effect of beaming could concentrate the soft emission, which would explain the softening of the emission in NGC 5408 X-1 that occurs when the source brightens as a result of its assumed precession \citep{Begelman:2006}.  We suggest, therefore, that the 115-day periodicity found by \cite{Strohmayer:2009} for NGC 5408 X-1 is more likely superorbital in nature and similar to SS~433.  This has implications for the determination of the black hole mass in ULXs, significantly reducing the mass needed to produce the observed properties.

\section{Conclusions}
If the nature of the ultra-luminous X-ray source (ULX) NGC 5408 X-1 is similar to that of the Galactic microquasar SS~433, then this ULX may be best described as an extragalactic high mass X-ray binary undergoing mass transfer at or near the Eddington rate, but at a lower inclination than would be typically seen in an accretion disc corona source \citep{Begelman:2006, Fabrika:2007, King:2009}.  This would mean that in the case of NGC 5408 X-1, we are viewing the hot inner accretion flow, substantially boosting the apparent brightness of this source.  The 115 day periodicity could be due to precession of the inner-disc/jet of NGC 5408 X-1, similar to SS~433, and not necessarily the orbital period as assumed in \cite{Strohmayer:2009}.
 
With {\it Chandra} or {\it XMM-Newton}, the best currently-available X-ray imaging telescopes, the count rate is typically too low to resolve the nature of the 115-day modulation in NGC 5408 X-1.  Settling this will be possible with a next-generation X-ray facility that has a greater collecting area.  Indeed, it is certainly possible that the entire class of ULXs are merely high mass X-ray binaries that are undergoing super-critical mass accretion onto a stellar-mass black hole, substantial beaming, or a combination of both.  If we can confirm that NGC 5408 X-1 does contain a stellar-mass black hole, the link between ULXs and intermediate-mass black holes will have been severely challenged.

\section*{Acknowledgments}
We are grateful to an anonymous referee for a number of helpful comments.  DLF acknowledges support from the US Department of Education in the form of a Graduate Assistance in Areas of National Need (GAANN) Fellowship (K. Stassun, PI), a fellowship from the Vanderbilt Initiative in Data-intensive Astrophysics (VIDA), the Vanderbilt--Cape Town Partnership Program, the International Academic Programmes Office (IAPO) at the University of Cape Town, and NSF grant AST-0849736.


\begin{thebibliography}{}
\bibitem[\protect\citeauthoryear{Bate et al.}{2000}]{Bate:2000} Bate M. R., Bonnell I. A.,  Clarke C. J., Lubow S. H., Ogilvie G. I., Pringle J. E., Tout C. A., 2000, \mnras, 317, 773
\bibitem[\protect\citeauthoryear{Begelman, King, \& Pringle}{Begelman et al.}{2006}]{Begelman:2006} Begelman M. C., King A. R., Pringle J. E., 2006, \mnras, 370, 399
\bibitem[\protect\citeauthoryear{Blundell, Bowler, \& Schmidtobreick}{Blundell et al.}{2008}]{Blundell:2008} Blundell K. M., Bowler M. G., Schmidtobreick L., 2008, \apjl, 678, L47
\bibitem[\protect\citeauthoryear{Charles \& Coe}{2006}]{Charles:2006} Charles P. A., Coe M. J., 2006, in Lewin W. H. G., van der Klis M., eds, Compact Stellar X-ray Sources. Cambridge University Press, Cambridge, p. 215
\bibitem[\protect\citeauthoryear{Charles et al.}{2008}]{Charles:2008} Charles P., Clarkson W., Cornelisse R., Shih I. C., 2008, New Astron. Rev., 51, 768
\bibitem[\protect\citeauthoryear{Crampton, Cowley, \& Hutchings}{Crampton et al.}{1980}]{Crampton:1980} Crampton D., Cowley A. P., Hutchings J. B., 1980, \apjl, 235, L131
\bibitem[\protect\citeauthoryear{Colbert et al.}{2004}]{Colbert:2004} Colbert E. J. M., Heckman T. M., Ptak A. F., Strickland D. K., 2004, \apj, 602, 231
\bibitem[\protect\citeauthoryear{Eikenberry et al.}{2001}]{Eikenberry:2001} Eikenberry S. S., Cameron P. B., Fierce B. W., Kull D. M., Dror D. H., Houck J. R., Margon B., 2001, \apj, 561, 1027
\bibitem[\protect\citeauthoryear{Fabbiano}{1989}]{Fabbiano:1989} Fabbiano G., 1989, \araa, 27, 87
\bibitem[\protect\citeauthoryear{Fabbiano \& White}{2006}]{Fabbiano:2006} Fabbiano G., White N. E., 2006, in Lewin W. H. G., van der Klis M., eds, Compact Stellar X-ray Sources. Cambridge University Press, Cambridge, p. 475
\bibitem[\protect\citeauthoryear{Fabbiano, Zezas, \& Murray}{Fabbiano et al.}{2001}]{Fabbiano:2001} Fabbiano G., Zezas A., Murray S. S., 2001, \apj, 554, 1035
\bibitem[\protect\citeauthoryear{Fabrika}{2004}]{Fabrika:2004} Fabrika S., 2004, ASPRv, 12, 1
\bibitem[\protect\citeauthoryear{Fabrika \& Abolmasov}{2007}]{Fabrika:2007} Fabrika~S., Abolmasov~P., 2007, in Kissler-Patig M., Walsh J. R., Roth M. M., eds, Science Perspectives for 3D Spectroscopy, Springer, p.309
\bibitem[\protect\citeauthoryear{Fabrika \& Mescheryakov}{2001}]{Fabrika:2001} Fabrika~S., Mescheryakov~A., 2001, in Schilizzi R. T., ed., Proc. IAU Symp. 205. Astron. Soc. Pac., San Francisco, p. 268
\bibitem[\protect\citeauthoryear{Frank, King, \& Lasota}{Frank et al.}{1987}]{Frank:1987} Frank J., King A. R., Lasota J.-P., 1987, \aap, 178, 137
\bibitem[\protect\citeauthoryear{Gies et. al}{2002}]{Gies:2002} Gies D. R., McSwain M. V., Riddle R. L., Wang Z., Wiita P. J., Wingert D. W., 2002, \apj, 566, 1069
\bibitem[\protect\citeauthoryear{Greene, Bailyn, \& Orosz}{Greene et al.}{2001}]{Greene:2001} Greene J., Bailyn C. D., Orosz J. A., 2001, \apj, 554, 1290
\bibitem[\protect\citeauthoryear{Greiner, Cuby, \& McCaughrean}{Greiner et al.}{2001}]{Greiner:2001} Greiner J., Cuby J. G., McCaughrean M. J., 2001, \nat, 414, 522
\bibitem[\protect\citeauthoryear{Hjellming \& Rupen}{1995}]{Hjellming:1995} Hjellming R. M., Rupen M. P., 1995, \nat, 375, 464
\bibitem[\protect\citeauthoryear{Kaaret \& Corbel}{2009}]{Kaaret:2009} Kaaret P., Corbel S, 2009, \apj, 697, 950
\bibitem[\protect\citeauthoryear{Kaaret et al.}{2003}]{Kaaret:2003} Kaaret P., Corbel S., Prestwich A. H., Zezas A., 2003, Science, 299, 365
\bibitem[\protect\citeauthoryear{Karachentsev et al.}{2002}]{K:2002} Karachentsev I. D., Sharina M.E., Dolphin A. E., Grebel E. K., Geisler D., Guhathakurta P., Hodges P. W., Karachentseva V. E., Sarajedini A., Seitzer P., 2002, \aap, 385, 21
\bibitem[\protect\citeauthoryear{Katz et al.}{1982}]{Katz:1982} Katz J. I., Anderson S. F., Margon B., Grandi S. A., 1982, \apj, 260, 780
\bibitem[\protect\citeauthoryear{King}{2002}]{King:2002} King A. R., 2002, \mnras, 335, L13
\bibitem[\protect\citeauthoryear{King}{2008a}]{King:2008a} King A. R., 2008a, \mnras, 385, L113
\bibitem[\protect\citeauthoryear{King}{2008b}]{King:2008b} King A., 2008b, New Astron. Rev., 51, 775
\bibitem[\protect\citeauthoryear{King}{2009}]{King:2009} King A. R., 2009, \mnras, 393, L41
\bibitem[\protect\citeauthoryear{King, Taam, \& Begelman}{King et al.}{2000}]{King:2000} King A. R., Taam R. E., Begelman M. C., 2000, \apjl, 530, L25
\bibitem[\protect\citeauthoryear{Kotani et al.}{2006}]{Kotani:2006} Kotani T., Trushkin S. A., Valiullin R., Kinugassa K., Safi-Harb S., Kawai N., Namiki M., 2006, \apj, 637, 486
\bibitem[\protect\citeauthoryear{Kotze \& Charles}{2010}]{Kotze:2010} Kotze M. M., Charles P. A., 2010, \mnras, 402, L16
\bibitem[\protect\citeauthoryear{Kozai}{1962}]{Kozai:1962} Kozai, Y., 1962, \aj, 67, 591
\bibitem[\protect\citeauthoryear{Kubota et al.}{2010}]{Kubota:2010} Kubota K., Ueda Y., Medvedev A., Barsukova E. A., Sholukhova O., Goranskij V. P., 2010, \apj, 709, 1374
\bibitem[\protect\citeauthoryear{Lang et al.}{2007}]{Lang:2007} Lang C. C., Kaaret P., Corbel S., Mercer A., 2007, \apj, 666, 79
\bibitem[\protect\citeauthoryear{Long et al.}{1981}]{Long:1981} Long K. S., d'Odorico S., Charles P. A., Dopita M. A., 1981, \apjl, 246, L61 
\bibitem[\protect\citeauthoryear{Makishima et al.}{2000}]{Makishima:2000} Makishima K., Kubota A., Mizuno T., Ohnishi T., Tashiro M., Asai K., Dotani T., Mitsuda K., Ueda Y., Uno S., Yamaoka K., Ebisawa K., Kohmura Y., Okada K., 2000, \apj, 535, 632
\bibitem[\protect\citeauthoryear{Maloney, Begelman, \& Nowak}{Maloney et al.}{1996}]{Maloney:1996} Maloney P. R., Begelman M. C., Nowak M. A., 1996, \apj, 472, 582
\bibitem[\protect\citeauthoryear{Maloney, Begelman, \& Nowak}{Maloney et al.}{1998}]{Maloney:1998} Maloney P. R., Begelman M. C., Nowak M. A., 1998, \apj, 504, 77
\bibitem[\protect\citeauthoryear{Margon}{1984}]{Margon:1984} Margon B., 1984, \araa, 22, 507
\bibitem[\protect\citeauthoryear{Mazeh et al.}{1980}]{Mazeh:1980} Mazeh T., Leibowitz E. N., Lahav O., Sheffer Y., 1980, IAUC, 3527, 2
\bibitem[\protect\citeauthoryear{Miller}{2005}]{Miller:2005} Miller, J. M., 2005, Ap\&SS, 300, 227
\bibitem[\protect\citeauthoryear{Miller \& Colbert}{2004}]{MandC:2004} Miller M. C., Colbert E. J. M., 2004, Int. J. Mod. Phys. D, 13, 1
\bibitem[\protect\citeauthoryear{Morgan, Remillard, \& Greiner}{Morgan et al.}{1997}]{Morgan:1997} Morgan E. H., Remillard R. A., Greiner J., 1997, \apj, 482, 993
\bibitem[\protect\citeauthoryear{Poutanen et al.}{2007}]{Poutanen:2007} Poutanen J., Lipunova G., Fabrika S., Butkevich A. G., Abolmasov P., 2007, \mnras, 377, 1187
\bibitem[\protect\citeauthoryear{Prestwich et al.}{2007}]{Prestwich:2007} Prestwich A. H., Kilgard R., Crowther P. A., Carpano S., Pollock A. M. T., Zezas A., Saar S. H., Roberts T. P., Ward M. J., 2007, \apjl, 669, L21
\bibitem[\protect\citeauthoryear{Rappaport, Podsiadlowski, \& Pfahl}{Rappaport et al.}{2005}]{Rappaport:2005} Rappaport S. A., Podsiadlowski P., Pfahl E., 2005, \mnras, 356, 401
\bibitem[\protect\citeauthoryear{Rau, Greiner, \& McCollough}{Rau et al.}{2003}]{Rau:2003} Rau A., Greiner J., McCollough M. L., 2003, \apjl, 590, L37
\bibitem[\protect\citeauthoryear{Remillard et al.}{1999}]{Remillard:1999} Remillard R. A., Morgan E. H., McClintock J. E., Bailyn C. D., Orosz J. A., 1999, \apj, 522, 397
\bibitem[\protect\citeauthoryear{Remillard et al.}{2002}]{Remillard:2002} Remillard R. A., Sobczak G. J., Muno M. P., McClintock J. E., 2002, \apj, 564, 962
\bibitem[\protect\citeauthoryear{Rico}{2008}]{Rico:2008} Rico J., 2008, \apjl, 683, L55
\bibitem[\protect\citeauthoryear{Shakura \& Sunyaev}{1973}]{Shakura:1973} Shakura N. I., Sunyaev R. A., 1973, \aap, 24, 337
\bibitem[\protect\citeauthoryear{Shaposhnikov \& Titarchuk}{2009}]{SandT:2009} Shaposhnikov N., Titarchuk L., 2009, \apj, 699, 453
\bibitem[\protect\citeauthoryear{Soria et al.}{2006}]{Soria:2006} Soria R., Fender R. P., Hannikainen D. C., Read A. M., Stevens I. R., 2006, \mnras, 368, 1527
\bibitem[\protect\citeauthoryear{Strohmayer}{2001}]{Strohmayer:2001} Strohmayer T. E., 2001, \apjl, 522, L49
\bibitem[\protect\citeauthoryear{Strohmayer}{2009}]{Strohmayer:2009} Strohmayer T. E., 2009, \apjl, 706, L210
\bibitem[\protect\citeauthoryear{Strohmayer \& Mushotzky}{2009}]{Strohmayer:2009b} Strohmayer T. E., Mushotzky R. F., 2009, \apj, 703, 1386
\bibitem[\protect\citeauthoryear{Strohmayer et al.}{2007}]{Strohmayer:2007} Strohmayer T. E., Mushotzky R. F., Winter L., Soria R., Uttley P., Cropper M., 2007, \apj, 660, 580
\bibitem[\protect\citeauthoryear{Titarchuk \& Fiorito}{2004}]{TF:2004} Titarchuk L. G., Fiorito R., 2004, \apj, 612, 988
\bibitem[\protect\citeauthoryear{Titarchuk, Lapidus, \& Muslimov}{Titarchuk et al.}{1998}]{TLM:1998} Titarchuk L., Lapidus I. I., Muslimov A., 1998, \apj, 499, 315
\bibitem[\protect\citeauthoryear{Vikhlinin et al.}{1994}]{Vikhlinin:1994} Vikhlinin A., Churazov E., Gilfanov M., Sunyaev R., Dyachkov A., Khavenson N., Kremnev R., Sukhanov K., 1994, \apj, 424, 395
\bibitem[\protect\citeauthoryear{Watson et al.}{1986}]{Watson:1986} Watson M. G., Stewart G. C., Brinkmann W., King A. R., 1986, \mnras, 222, 261
\bibitem[\protect\citeauthoryear{Wen et al.}{2006}]{Wen:2006} Wen L., Levine A. M., Corbet R. H. D., Bradt H. V., 2006, \apjs, 163, 372
\bibitem[\protect\citeauthoryear{Zampieri \& Roberts}{2009}]{Zampieri:2009} Zampieri L., Roberts T. P., 2009, \mnras, 400, 677

\end{thebibliography}
\end{document}